\documentclass[pdftex,twocolumn]{aastex63}

\usepackage{apjfonts,graphicx,graphics}

\usepackage{hyperref}
\usepackage{graphicx}
\usepackage{mathptmx}
\usepackage{amsmath}
\usepackage{wasysym}
\usepackage{times}
\usepackage{epsfig}
\usepackage{ulem}
\usepackage{epstopdf}
\usepackage{appendix}
\usepackage{units}
\usepackage{import}
\newbox{\bigpicturebox}

\usepackage[capitalize]{cleveref}
\usepackage{CJKutf8}

\usepackage{amssymb}

\newcommand{\chinesenameXiaoweiduan}{{\begin{CJK}{UTF8}{gbsn}(段晓苇)\end{CJK}}}

\newcommand{\chinesenameXiaodianchen}{{\begin{CJK}{UTF8}{gbsn}(陈孝钿)\end{CJK}}}

\newcommand{\chinesenameLicaideng}{{\begin{CJK}{UTF8}{gbsn}(邓李才)\end{CJK}}}

\newcommand{\chinesenameHuaweizhang}{{\begin{CJK}{UTF8}{gbsn}(张华伟)\end{CJK}}}

\newcommand{\chinesenameWeijiasun}{{\begin{CJK}{UTF8}{gbsn}(孙唯佳)\end{CJK}}}

\renewcommand{\thefootnote}{*}

\newcommand{\m}{\rm\; m}

\newcommand{\s}{\rm\; s}

\newcommand{\K}{\rm\; K}

\newcommand{\g}{\rm\; g}
\newcommand{\W}{\rm\; W}
\begin{document}

\title{Further evidence of shocks in the first-overtone RR Lyrae pulsators: first detection of shock-triggered magnesium emissions}


\author[0000-0002-6573-6719]{Xiao-Wei Duan \chinesenameXiaoweiduan}\footnote{\href{mailto:duanxw@pku.edu.cn}{duanxw@pku.edu.cn}, \href{mailto:licai@bao.ac.cn}{licai@bao.ac.cn}, \href{mailto:zhanghw@pku.edu.cn}{zhanghw@pku.edu.cn}}

\affiliation{Department of Astronomy, Peking University, Yi He Yuan Road 5, Hai
Dian District, Beijing 100871, China}
\affiliation{Kavli Institute for Astronomy \& Astrophysics, Peking University,
Yi He Yuan Road 5, Hai Dian District, Beijing 100871, China}

\author[0000-0002-3279-0233]{Weijia Sun \chinesenameWeijiasun}
\affiliation{Department of Astronomy, Peking University, Yi He Yuan Road 5, Hai
Dian District, Beijing 100871, China}
\affiliation{Kavli Institute for Astronomy \& Astrophysics, Peking University,
Yi He Yuan Road 5, Hai Dian District, Beijing 100871, China}

\author[0000-0001-7084-0484]{Xiaodian Chen \chinesenameXiaodianchen}
\affiliation{CAS Key Laboratory of Optical Astronomy, National Astronomical
Observatories, Chinese Academy of Sciences, Beijing 100101, China}
\affiliation{School of Astronomy and Space Science, University of the Chinese
Academy of Sciences, Huairou 101408, China}
\affiliation{Department of Astronomy, China West Normal University, Nanchong
637009, China}

\author[0000-0001-9073-9914]{Licai Deng \chinesenameLicaideng}
\affiliation{CAS Key Laboratory of Optical Astronomy, National Astronomical
Observatories, Chinese Academy of Sciences, Beijing 100101, China}
\affiliation{School of Astronomy and Space Science, University of the Chinese
Academy of Sciences, Huairou 101408, China}
\affiliation{Department of Astronomy, Peking University, Yi He Yuan Road 5, Hai
Dian District, Beijing 100871, China}
\affiliation{Department of Astronomy, China West Normal University, Nanchong
637009, China}

\author[0000-0002-7727-1699]{Huawei Zhang \chinesenameHuaweizhang}

\affiliation{Department of Astronomy, Peking University, Yi He Yuan Road 5, Hai
Dian District, Beijing 100871, China}
\affiliation{Kavli Institute for Astronomy \& Astrophysics, Peking University,
Yi He Yuan Road 5, Hai Dian District, Beijing 100871, China}

\begin{abstract}
The behavior of the shock wave in the atmosphere of the non-fundamental mode RR Lyrae pulsator remains a mystery. In this work, we firstly report a blueshifted Mg triplet emission in continuous spectroscopic observations for a non-Blazhko RRc pulsator (Catalina-1104058050978) with LAMOST medium resolution spectra. We analyse the photometric observations from Catalina Sky Survey of this RRc pulsator with pre-whitening sequence method and provide the ephemeris and phases. An additional frequency signal with $P_1/P_x = 0.69841$ is detected and discussed. The redshift and radial velocity of the spectra are provided by fitting process with $S\acute{e}rsic$ functions and cross-correlation method. Moreover, we plot the variation of H$\alpha$ and Mg lines in a system comoving with the pulsation. Clear evolution of comoving blueshifted hydrogen and Mg emission is observed, which further confirms the existence of shock waves in RRc pulsators. The shock-triggered emission lasts over $15\%$ of the pulsation cycle, which is much longer than the previous observations.
\end{abstract}


\section{Introduction}\label{sec:intro}

RR Lyrae stars are classified as fundamental mode (RRab), the first-overtone mode (RRc), or double-mode (RRd) RR Lyrae variables according to the various number of oscillation modes \citep{Soszy2011AcA....61....1S}. The most intriguing problem of RR Lyrae pulsators is called ``Blazhko effect'' \citep{Bla1907AN....175..325B,Kolenberg2006A&A...459..577K}. It means some of RR Lyrae variables display periodic amplitude and/or phase modulations overtime, which leads to miscalculation of the ephemeris and limits the use of RR Lyrae variables as ``standard candles'' for precise distance determinations \citep{Longmore1986MNRAS.220..279L,Catelan2004ApJS..154..633C}. Up to now, the physical mechanism of the Blazhko effect is still under debate.

\cite{Chadid2014AJ....148...88C} proposed that the Blazhko effect is triggered by a dynamical interaction between a structure with multi-shock and an outflowing wind in a coronal structure. To recognize shock waves in RR Lyrae stars, hydrogen emission lines on the spectra are considered as the most important tracers, which contain three main types, named ``three apparitions'' \citep{Preston2011AJ....141....6P}. The ``first apparition'' is a blueshifted hydrogen emission, generated by the main shock before maximum luminosity \citep{Sanford1949}.  The ``second apparition'' is a blueshifted weak hump, which appears near $\phi=0.7$ \citep{Gillet1988}. It is suggested to be generated by photospheric compression from the collision between the inner deep expanding atmosphere and the outer layers during a ballistic infalling motion \citep{Hill1972}. The ``third apparition'' is a weak redshifted bump that appears near $\phi=0.3$ \citep{Preston2011AJ....141....6P}. \cite{Gillet2017A&A...607A..51G} represented the first observation of the ``third apparition'' and showed the evidence of a supersonic infalling motion process of the atmosphere.

The Blazhko effect is a frequently observed on the light curves of RRab and RRc variables. Shock wave signatures are systemically observed and analysed in RRab stars \citep{Chadid1996A&A...308..481C,Gillet2019}. However, the existence of shock wave in the atmosphere of RRc pulsators has not been thoroughly proved. Owing to limited spectroscopic observation, the dynamical evolution of the atmosphere of RRc pulsators remains a mystery. \cite{Duan2021ApJ...918....3D} conducted searching of the ``first apparition'' in SDSS and LAMOST low-resolution spectroscopic data based with one-dimensional pattern recognition algorithm. \cite{Duan2021ApJ...909...25D} reported possible detection of the ``first apparition'' in non-fundamental mode RR Lyrae pulsators, which indicates the possible existence of shock wave in RRc and RRd pulsators. Meanwhile, \cite{Benk2021MNRAS.500.2554B} presented the first spectroscopic time series observations on one of the brightest northern RRc stars, T Sex. They reported a periodic distortion of the H$\alpha$ line profile in the pulsation cycle and suggested that it is probably caused by the turbulent convection. They also presented an emission on the Na D profile at $\phi = 0.438$, which possibly indicates a weak shock wave. More observations, especially time series, are needed to unveil the shock wave in non-fundamental mode RR Lyrae pulsators.

In this work, we present time-series spectroscopical observation of a non-Blazhko RRc pulsator, Catalina-1104058050978, with Large Sky Area Multi-Object Fiber Spectroscopic Telescope Medium Resolution Survey \citep[LAMOST-MRS,][]{Liu2020arXiv200507210L} single-epoch spectra. We choose a non-Blazhko RRc variable because the phase can be accurately determined and the causality between spectroscopic features and shock wave can be proved. We firstly report the shock-triggered magnesium emission in a RRc pulsator, which further confirms the existence of shock waves in the atmosphere of the non-fundamental RR Lyrae stars and also influences the understanding of the origin of the Blazhko effect.

\section{Observations}\label{section:observations}

The spectroscopic observation is collected from Large sky Area Multi-Object fiber Spectroscopic Telescope \citep[LAMOST,][]{Deng2012RAA....12..735D,Zhao2012,Cui2012} DR1-9 median-resolution and single-epoch spectra. LAMOST is a Chinese national scientific research facility, consisting of the LAMOST ExtraGAlactic Survey (LEGAS) and the LAMOST Experiment for Galactic Understanding and Exploration (LEGUE). It scans millions of objects in the northern sky and surveys a large volume of space in high efficiency.

Following the low-resolution survey (LRS) with $R \sim 1800$ \citep{Luo2015RAA....15.1095L}, a medium-resolution survey (MRS) with $R \sim 7500$ has been conducting begun 2017. Since LAMOST DR6, LAMOST-MRS spectra have been successfully applied to stellar physics \citep{Zong2020ApJS..251...15Z,Sun2021arXiv210801213S,Sun2021arXiv210801212S}. For LAMOST-MRS spectra, each plate is observed 3 to 6 times. The exposure time is usually 20 minutes each. Single-epoch spectra give us the chance to avoid overwhelming of features owing to co-addition of spectra. The red camera of each spectrograph of LAMOST-MRS covers the wavelength from 6300 to 6800 $\rm \mathring{A}$, including the range of H$\alpha$ lines. The blue camera covers the wavelength from 4950 to 5350 $\rm \mathring{A}$, including the range of Mg triplet. In this work, the spectra were observed from January 2019 to April 2021.

The photometric time series is got from the Catalina Sky Survey, which began in 2004 \citep{Drake2009ApJ...696..870D}. Three telescopes are used to scan most of the sky in the range of $-75^\circ<\delta<+65^\circ$ since 2003. The sub-surveys are the Catalina Schmidt Survey (CSS), the Mount
Lemmon Survey (MLS), and the Siding Spring Survey (SSS). Most of the observations have been taken as four images. Every field will be repeatedly observed after 10 minutes. The images are unfiltered with aperture photometry carried out by the $\rm SE_{\rm XTRACTOR}$ program \citep{Bertin1996A&AS..117..393B}. The photometric time series and analyses of RR Lyrae pulsators have been published by \cite{Drake2013ApJ...763...32D,Drake2013bApJ...765..154D,Torrealba2015MNRAS.446.2251T,Drake2014ApJS..213....9D,Drake2017MNRAS.469.3688D}.

\section{Analysis of the photometric observations}\label{section:photometry}

We analyze $V$-band light curve of this RRc pulsator from Catalina Sky Survey (ID: 1104058050978, hereafter: Catalina-1104058050978) with R.A.(J2000) = $161.11333$ deg, Decl.(J2000) = $+5.22253$ deg, $V = 14.02$ mag, with a time span of 3127.25 days by a standard successive pre-whitening method \citep{Moskalik2009MNRAS.394.1649M}. The photometric observational points is fitted by the non-linear least-square procedure with the sine series of this following form at each step: 
\begin{equation}
    m(t) = m_0 + \sum^N_{k=1} A_k {\rm sin}(2\pi f_k t+\phi_k),
	\label{eq:fulllightcurve}
\end{equation}
where $f_k$ represents independent frequencies detected in discrete Fourier transform of the photometric data and their possible linear combinations.

\begin{figure}[tb]
\begin{center}
\begin{tabular}{c}
\includegraphics*[width=90mm,height=6cm]{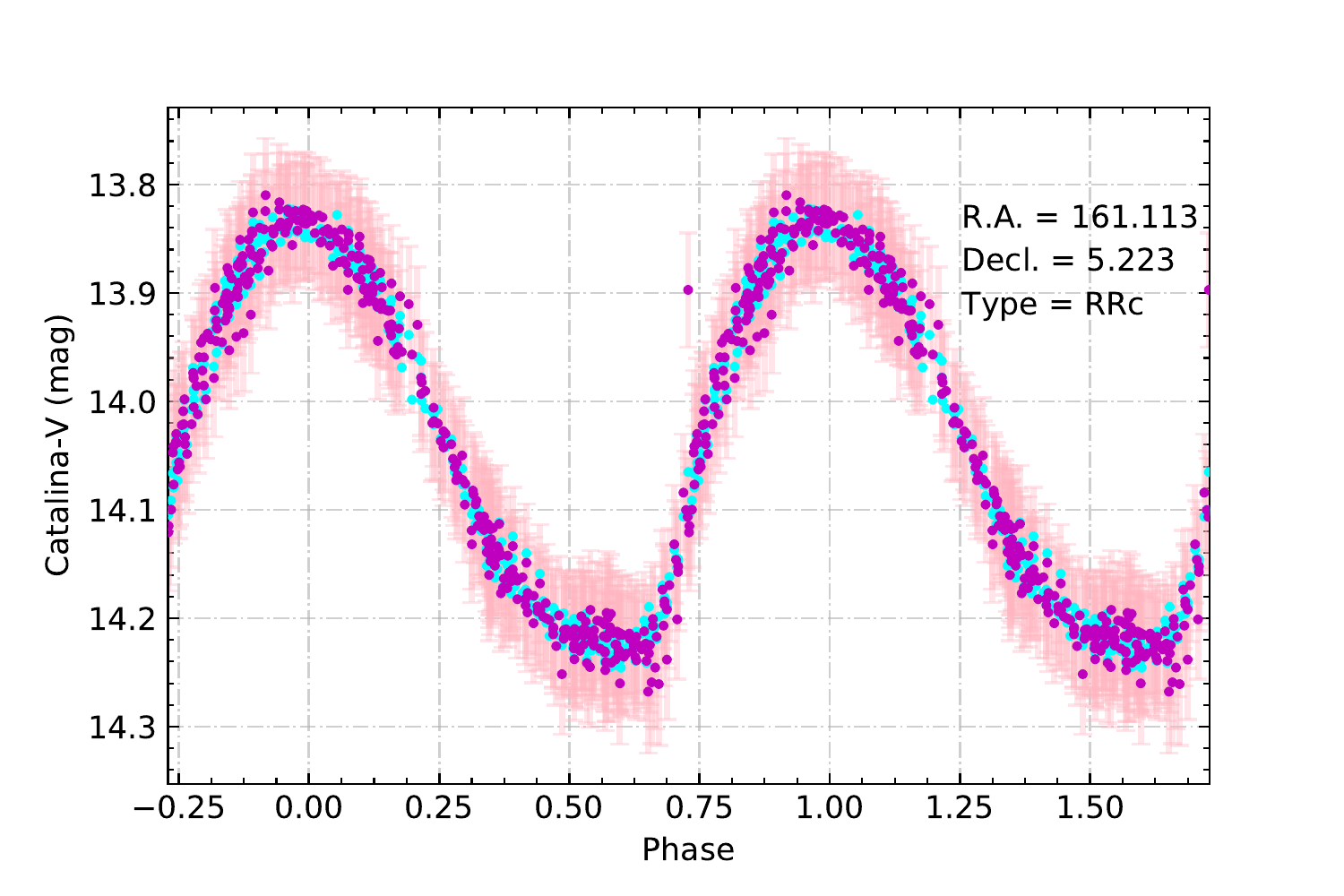}
\end{tabular}
\caption{\label{fig:newlightcurve} Regeneration of light curve of RRc pulsator Catalina-1104058050978 with $full$ $light$ $curve$ $solution$. Magenta points and light pink error bars are photometric observations from Catalina-V. Cyan points are regenerated points from the $full$ $light$ $curve$ $solution$.}
\end{center}
\end{figure}

The residuals of the fitting processes are utilized to search for frequencies in the next step. Afterwards, a new Fourier series containing all frequencies detected so far are fitted to the photometric time series again. The process repeats until no new significant frequency signal is found and the residuals become virtually white noise. The frequencies are regarded as unresolved if $\Delta < 2/T$, where T denotes the length of the photometric time series. Here $2/T \approx 0.0006$. We parameterize the light curve with the detected frequency signals and build a $full$ $light$ $curve$ $solution$ for this RRc pulsator. The fitting result of the light curve is shown in Figure~\ref{fig:newlightcurve}. According to the first-overtone frequency $f_1 = 3.03514$, we derive the first-overtone period $P_{1} = 0.32947
d$ and $A_1 = 0.19926$ mag. The signals detected in the pre-whitening sequence should be $f_1$, $nf_1$ (linear combinations of $f_1$), and $nf_1+\lambda_cf_c$,
where $f_c\approx1$. $f_c$ can be explained as the daily cadence. The parameter of Catalina-1104058050978 is summarized in Table~\ref{table:parameterRRc}.

\begin{table*}
\begin{center}
\caption{\label{table:parameterRRc} Parameters of RRc pulsator Catalina-1104058050978.}
\setlength{\tabcolsep}{1mm}{
\begin{tabular}{lccccccccc}
\hline
Type   & R.A.(J2000) & Decl.(J2000) & $V$ & $P_{1}$ & $P_x$ &   $P_{1}/P_x$   & $A_1$  & $A_x$ & Ephemeris (MJM)\\
$  $          & ${}^{\circ}$ & ${}^{\circ}$ & mag & day & day &   & mag & mag & min \\
\hline
RRc  & $161.11333$ & $+5.22253$ & $14.02$ & $0.32947$ & $0.23011$ & $0.69841$ & $0.19926$ & $0.01305$ & $84234924.67416$\\
\hline
\end{tabular}}
\end{center}
\tablecomments{\\\hspace{\textwidth}
	1. $P_1$ denotes the first-overtone period, while $A_1$ denotes the first-overtone amplitude. \\\hspace{\textwidth}
	2. $P_x$ represents the additional period, while $A_x$ represents the additional amplitude. \\\hspace{\textwidth}
	3. Ephemeris indicates the location of maximum luminosity, which is in unit of minute.}
\end{table*}

Apart from the first-overtone frequency, an additional frequency $f_x = 4.34579$ is also detected, with $P_1/P_x = 0.69841$. Many RR Lyrae variables show additional signals with small-amplitude, which cannot be explained as radial modes or Blazhko effects \citep{Moskalik2013ASSP...31..103M}. \cite{Olech2009A&A...494L..17O} firstly reported two RRc variables in the globular cluster $\omega$ Cen with an additional mode beyond first-overtone mode with $P_x/P_1 = 0.612-0.614$, and four other RRc stars with period ratios of $\sim 0.61$, depending on the choice of aliases. \cite{Soszynski2009AcA....59....1S} provided two RRc pulsators from OGLE with $P_x/P_1 = 0.603-0.610$. \cite{Suveges2012MNRAS.424.2528S} reported two RRc variables with $P_x/P_1 = 0.607-0.624$ from SDSS Stripe. \cite{Szabo2014A&A...570A.100S} revisited CoRoT RR Lyrae variables, including two RRc stars with $P_x/P_1 = 0.613-0.615$.

\cite{Moskalik2015MNRAS.447.2348M} presented photometric analysis of four RRc stars observed with the Kepler space telescope. All four stars are found with an additional modes with $P_x/P_1 = 0.612-0.632$. \cite{Molnar2015MNRAS.452.4283M} presented a detailed analysis of 33 RR Lyrae variables in Pisces observed with the $Kepler$ space telescope over the K2 Two-Wheel Concept Engineering Test and provide three RRc stars with $P_x/P_1 = 0.603-0.620$. \cite{Sodor2017MNRAS.465L...1S} discovered a new $Kepler$ RRc pulsator, KIC 2831097, with $P_x/P_1 = 0.612$. \cite{Benk2021MNRAS.500.2554B} provided spectroscopic and photometric time series of RRc pulsator T Sex, with $P_x/P_1 = 0.643$. \cite{Dziembowski2016CoKon.105...23D} suggested that additional ratio with $P_x/P_1 = 0.60-0.64$ is a harmonic of the non-radial mode of moderate degree ($l = 8,9$).

\begin{figure}
\begin{center}
\begin{tabular}{c}
\includegraphics*[width=90mm,height=6cm]{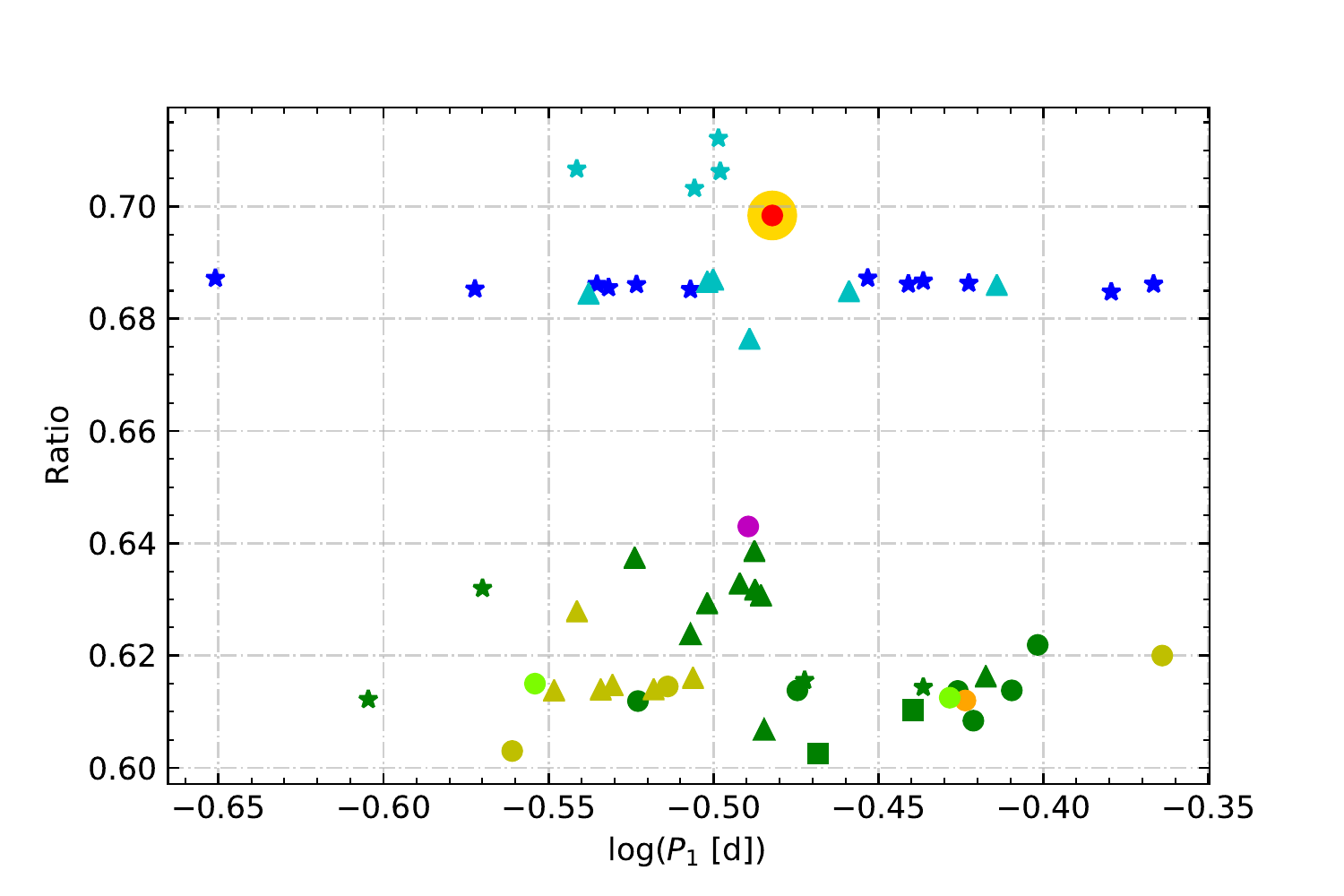}
\end{tabular}
\caption{\label{fig:petersendiagram} Petersen diagram for RRc pulsators. Corresponding relation: green circle - Olech \& Moskalik (2009), green square - Soszy{\'n}ski et al. (2009), green triangle - S{\"u}veges et al. (2012), law green circle - {Szab{\'o}} et al. (2014), green star - Moskalik et al. (2015), blue star - Netzel et al. (2015), yellow circle - {Moln{\'a}r} et al. (2015), orange circle - S$\acute{\rm o}$dor et al. (2017), cyan star - Prudil et al. (2017), yellow triangle - $RR_{0.61}$ stars in Netzel \& Smolec (2019), cyan triangle - $RR_{0.68}$ stars in Netzel \& Smolec (2019), green triangle - stars with $0.5f_x$ detected in Netzel \& Smolec (2019), magenta circle -  Benk$\H{o}$ et al. (2021), red circle with gold edge - Catalina-1104058050978 in this work.}
\end{center}
\end{figure}

\cite{Netzel2015MNRAS.451L..25N} reported a new group of double-periodic RR Lyrae variables in the Optical Gravitational Lensing Experiment $-$ IV (OGLE-IV) Galactic bulge photometry, including 11 RRc pulsators with $P_1/P_x$ ranging from $0.685$ to $0.687$. They also reported one RRc star from Kepler with $P_1/P_x = 0.68672$. \cite{Prudil2017MNRAS.465.4074P} discovered four RRc pulsators with $P_1/P_x$ ranging from $0.68-0.72$. \cite{Netzel2019MNRAS.487.5584N} analysed photometry observation of the collection of RRc and RRd pulsators towards OGLE. They suggested that RR Lyrae variables with additional small-amplitude signals can be divided into two groups: $RR_{0.61}$ and $RR_{0.68}$. In $RR_{0.61}$ group, the period of additional signal is shorter than the first-overtone period. The ratio is defined as $P_x/P_1$. While in $RR_{0.68}$, the reverse applies. Petersen diagram is shown in Figure~\ref{fig:petersendiagram}.

Moreover, the folded light curve of Catalina-1104058050978 does not show long-term modulation, which suggests that this RRc pulsator is a non-Blazhko star. The analysis for photometric and spectroscopic time series of RR Lyrae pulsators with Blazhko effect are required to be conducted at the same or similar time. Because long-term modulations make the phases of the spectra can not be determined accurately. On the contrary, researches for RR Lyrae variables without the Blazhko effect will not be restricted by this problem. 
The ephemeris denotes the location of the maximum luminosity. We determine the ephemeris as $84234924.67416$ min (MJM) in this work.

\section{Radial velocity}\label{sec:rv}

Two methods are used to derive radial velocity of the spectra. One is spectral fitting method of H$\alpha$ lines. One is data-driven method SLAM, based on the cross-correlation method.

\subsection{Spectral fitting method}\label{sec:fitting}

The absorption components of H$\alpha$ lines are fitted by the $scale$ $width$ $versus$ $shape$ method \citep{Sersic1968adga.book.....S,Clewley2002MNRAS.337...87C}. $S\acute{e}rsic$ functions \citep{Xue2008} as:
\begin{eqnarray}\label{eq:Sersicprofile}
y =m - a e^{-(\frac{\left|{\lambda_{abs}}-{\lambda_{0abs}}\right|}{b})^c},
\end{eqnarray}
is used to fit the profile. Uncertainties are produced by error propagation with the covariance matrix and Monte Carlo method \citep{Andrae2010arXiv1009.2755A}. 

The radial velocity of the blueshifted H$\alpha$ emission in the stellar rest
frame is provided as:
\begin{eqnarray}\label{eq:Vshock}
V_{\rm e1,\alpha} = c\frac{(\lambda_{\rm e1,\alpha}-\lambda_{\rm
ab})}{\lambda_{0}},
\end{eqnarray}
where $\lambda_{\rm e1,\alpha}$ represents the central wavelength of the emission line. $\lambda_{\rm ab}$ represents the central
wavelength of the absorption line. $\lambda_{0}$ represents the laboratory
wavelength. The results of $V_r$ from the fitting process is displayed in Table~\ref{table:parameterrv} as $V_{rf}$.

\subsection{Data-driven method: SLAM}\label{sec:slam}

The Stellar LAbel Machine (SLAM) is a data-driven method based on support vector regression (SVR), reported by \cite{Zhang2020ApJS..246....9Z}. It can estimate precise stellar labels in LAMOST DR5 with high efficiency. SLAM uses a generative model which adjusts the model complexity automatically and fetches information from the spectra for a wide range of spectral types. SLAM shows good performance on application to the LAMOST medium-resolution survey \citep{Zhang2020RAA....20...51Z}. After normalization, the radial velocity $V_r$ is estimated using the cross-correlation function method, respect to a collection of synthetic spectra with the ATLAS9 model atmospheres \citep{AllendePrieto2018A&A...618A..25A}. The results of $V_r$ from SLAM is displayed in Table~\ref{table:parameterrv} as $V_{rs}$. 

\begin{figure}[tb]
\begin{center}
\begin{tabular}{c}
\includegraphics*[width=90mm,height=6cm]{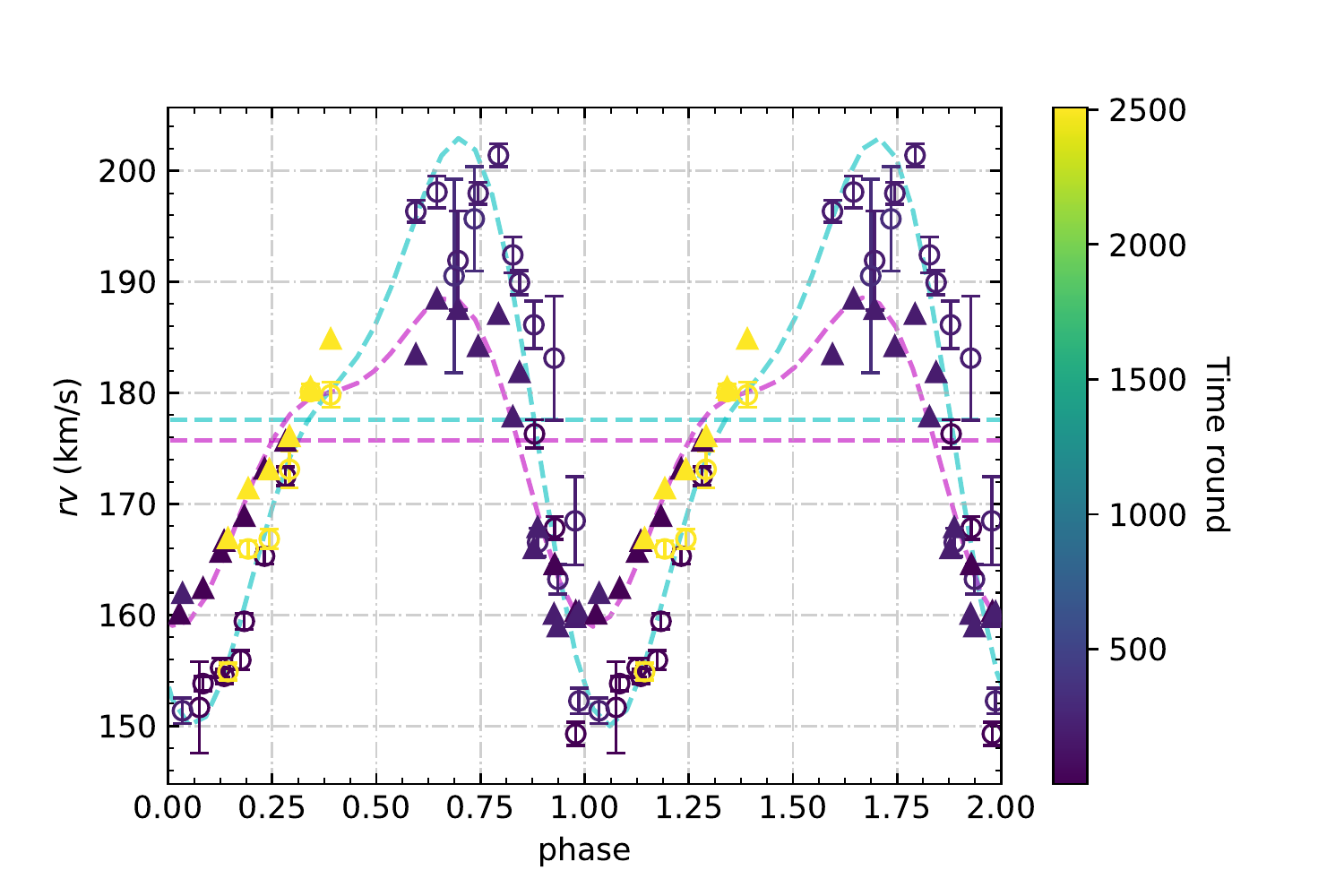}
\end{tabular}
\caption{\label{fig:radialvelocity} The radial velocity curve for LAMOST median-resolution spectroscopic time series of Catalina-1104058050978. Corresponding relation: circle - $V_r$ from fitting process with $S\acute{e}rsic$ function, triangle - $V_r$ from SLAM. The variation of color indicates the evolution of time.}
\end{center}
\end{figure}

Comparing to spectral fitting method, SLAM method only uses the blue segment of spectrum to avoid probable emission contamination, while spectral fitting method considers H$\alpha$ profiles. The result of $V_r$ from fitting process with $S\acute{e}rsic$ function and SLAM are shown in Figure~\ref{fig:radialvelocity}.  The amplitude of the result from fitting process is larger than that from SLAM. We fit the radial velocity by a second-order Fourier function with $P = 1$ and determine the mean radial velocity as $V_{mf} = 177.61$ km/s for fitting process and $V_{ms} = 175.71$ km/s for SLAM. 

\subsection{Fitting process for blueshifted hydrogen emission}

Some of the spectra display blueshifted hydrogen emission. For this case, we use two
$S\acute{e}rsic$ functions \citep{Yang2014,Duan2020CoBAO..67..181D} as:
\begin{eqnarray}\label{eq:Sersicprofile}
y =m - a e^{-(\frac{\left|{\lambda_{abs}}-{\lambda_{0abs}}\right|}{b})^c} + f e^{-(\frac{\left|{\lambda_{e}}-{\lambda_{0e}}\right|}{g})^h},
\end{eqnarray}
to fit the profile. Uncertainties are also calculated by error propagation with the covariance matrix and Monte Carlo method. 

\begin{figure*}[tb]
\begin{center}
\begin{tabular}{c}
\includegraphics*[width=90mm,height=20.2cm]{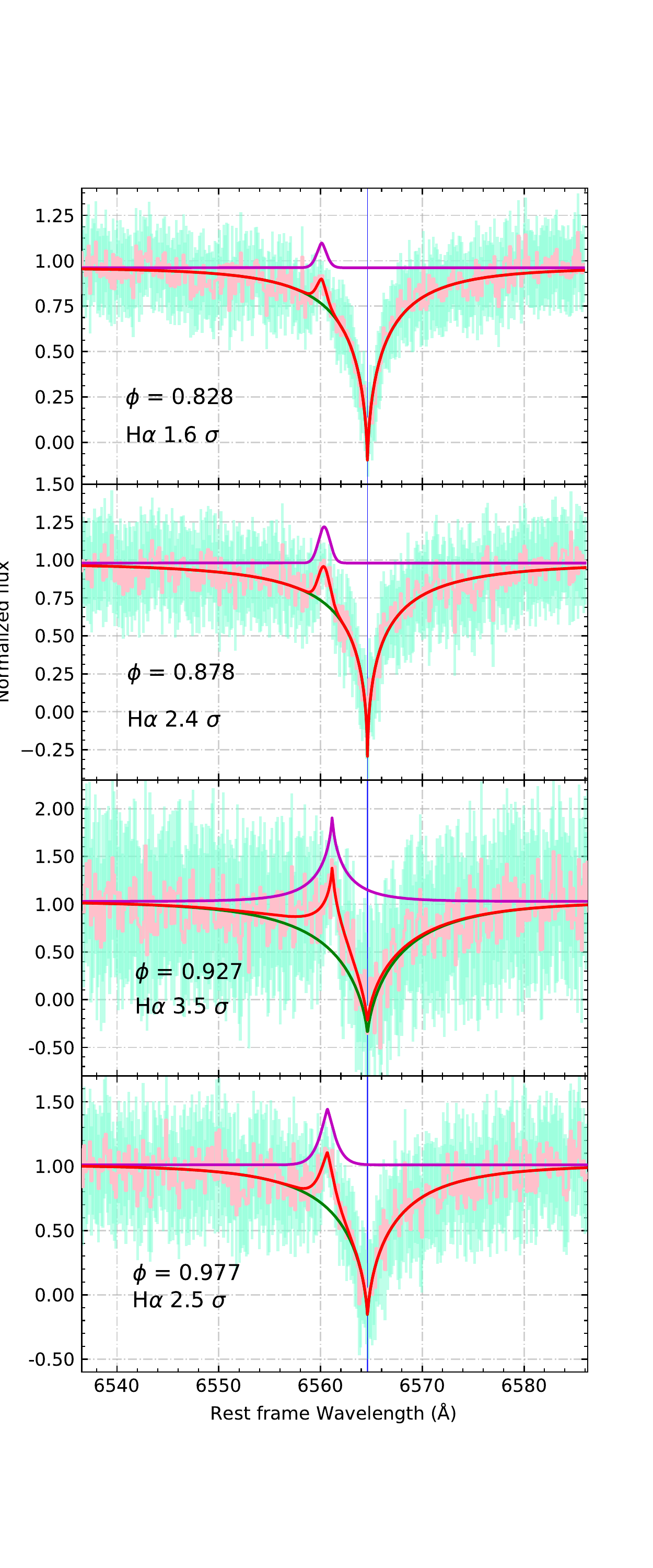}
\end{tabular}
\caption{\label{fig:blueshiftfitting} Visualization of the fitting results
of the blueshifted H$\alpha$ emission in Catalina-1104058050978. The wavelength axis is shown in the
stellar rest frame. The fitted blueshifted H$\alpha$ emission lines are displayed
as magenta profiles. Green profiles represent fitted absorption lines. Red profiles
denote the combination of fitted emission and absorption profiles. Vertical
blue lines represent the H$\alpha$ laboratory wavelength. The significances of the emission profiles
are calculated by the signal-to-noise ratio.}
\end{center}
\end{figure*}

As for the ``first apparitions'', the redshift in the frame of the observer $z_{\rm
e1,\alpha}$, the radial velocity in the stellar rest frame $V_{\rm e1,\alpha}$,
the normalized flux Flux$_{\rm e1,\alpha}$ and FWHM of
the emission and absorption FWHM$_{\rm e1,\alpha}$ in the stellar rest frame are displayed in Table~\ref{table:parameterrv}. The fitting process is visualized in Figure~\ref{fig:blueshiftfitting}.

\begin{table*}[tbp]
\begin{center}
\caption{\label{table:parameterrv} Radial velocity curve of Catalina-1104058050978 from LAMOST.}
\setlength{\tabcolsep}{1mm}{
\begin{tabular}{lccccccccc}
\hline
Phase  & $z_{fit}$ & $V_{rf}$ & $z_{SLAM}$ & $V_{rs}$  & $z_{\rm
e1,\alpha}$ & $V_{\rm e1,\alpha}$ &  Flux$_{\rm e1,\alpha}$ & FWHM$_{\rm
e1,\alpha}$\\
$  $          & $ $ & km/s & $ $ & km/s   & $ $ &
km/s &   & $\rm \mathring{A}$   \\
\hline
0.880  & $
5.88E-4\pm4.24E-6$ & $176.31\pm1.27$ & $/$ & $/$  &  &  &  &  \\
0.928  & $
5.60E-4\pm3.47E-6$ & $167.84\pm1.04$ & $
5.49E-4$ & $164.57$  &  &  &  &  \\
0.979  & $
4.98E-4\pm3.47E-6$ & $149.29\pm1.04$ & $
5.35E-4$ & $160.33$  &  &  &  &  \\
\hline
+1  &  &  &  &   &  &  &  &  \\
\hline
0.027  & $
4.32E-4\pm3.61E-5$ & $129.46^{*}\pm10.81^{*}$ & $
5.34E-4$ & $160.11$  &  &  &  &  \\
0.076  & $
5.06E-4\pm1.38E-5$ & $151.67\pm4.13$ & $/$ & $/$  &  &  &  &  \\
0.126  & $
5.18E-4\pm2.91E-6$ & $155.23\pm0.87$ & $5.53E-4$ & $165.68$  &  &  &  &  \\
0.175  & $
5.20E-4\pm2.93E-6$ & $155.94\pm0.88$ & $/$ & $/$  &  &  &  &  \\
\hline
+10  &  &  &  &   &  &  &  &  \\
\hline
0.084  & $
5.13E-4\pm2.13E-6$ & $153.81\pm0.64$ & $5.42E-4$ & $162.41$  &  &  &  &  \\
0.135  & $
5.15E-4\pm2.27E-6$ & $154.47\pm0.68$ & $5.56E-4$ & $166.67$  &  &  &  &  \\
0.183  & $
5.32E-4\pm2.40E-6$ & $159.43\pm0.72$ & $5.63E-4$ & $168.90$  &  &  &  &  \\
0.232  & $
5.51E-4\pm2.37E-6$ & $165.30\pm0.71$ & $5.78E-4$ & $173.16$  &  &  &  &  \\
0.282  & $
5.75E-4\pm2.84E-6$ & $172.49\pm0.85$ & $5.86E-4$ & $175.68$  &  &  &  &  \\
\hline
+188  &  &  &  &   &  &  &  &  \\
\hline
0.595  & $
6.55E-4\pm3.27E-6$ & $196.36\pm0.98$ & $6.12E-4$ & $183.48$  &  &  &  &  \\
0.645  & $
6.61E-4\pm4.80E-6$ & $198.10\pm1.44$ & $6.29E-4$ & $188.48$  &  &  &  &  \\
0.696  & $
6.40E-4\pm1.48E-5$ & $191.93\pm4.45$ & $6.26E-4$ & $187.55$  &  &  &  &  \\
0.745  & $
6.60E-4\pm3.27E-6$ & $197.99\pm0.98$ & $6.14E-4$ & $184.21$  &  &  &  &  \\
0.793  & $
6.72E-4\pm3.45E-6$ & $201.41\pm1.03$ & $6.24E-4$ & $187.11$  &  &  &  &  \\
0.844  & $
6.34E-4\pm3.66E-6$ & $189.94\pm1.10$ & $6.07E-4$ & $181.86$  &  &  &  &  \\
\hline
+203  &  &  &  &   &  &  &  &  \\
\hline
0.828  & $
6.38E-4\pm4.57E-6$ & $191.16\pm1.37$ & $5.93E-4^{*}$ & $177.86^{*}$  & $
-4.84E-5\pm1.25E-5$ & $-205.53\pm4.84$ & $0.14$ & $1.09$ \\
0.878  & $
6.14E-4\pm4.77E-6$ & $184.00\pm1.43$ & $5.54E-4^{*}$ & $166.00^{*}$  & $
-3.49E-5\pm1.02E-5$ & $-194.35\pm4.33$ & $0.24$ & $1.34$ \\
0.927  & $
4.95E-4\pm9.08E-6$ & $148.37\pm2.72$ & $5.34E-4^{*}$ & $160.08^{*}$  & $
-3.39E-5\pm7.83E-6$ & $-158.50\pm3.84$ & $0.91$ & $1.50$ \\
0.977  & $
5.31E-4\pm1.02E-5$ & $159.28\pm3.07$ & $5.33E-4^{*}$ & $159.81^{*}$  & $
-6.88E-5\pm1.04E-5$ & $-179.81\pm4.37$ & $0.44$ & $1.54$ \\
\hline
+209  &  &  &  &   &  &  &  &  \\
\hline
0.887  & $
5.55E-4\pm4.24E-6$ & $166.53\pm1.27$ & $5.60E-4$ & $167.90$  &  &  &  &  \\
0.936  & $
5.44E-4\pm4.42E-6$ & $163.21\pm1.33$ & $5.30E-4$ & $158.96$  &  &  &  &  \\
0.986  & $
5.08E-4\pm3.86E-6$ & $152.25\pm1.16$ & $5.35E-4$ & $160.28$  &  &  &  &  \\
\hline
+210  &  &  &  &   &  &  &  &  \\
\hline
0.035  & $
5.05E-4\pm3.85E-6$ & $151.37\pm1.15$ & $5.40E-4$ & $161.96$  &  &  &  &  \\
\hline
+309  &  &  &  &   &  &  &  &  \\
\hline
0.687  & $
6.36E-4\pm2.91E-5$ & $190.54\pm8.73$ &   &    &  &  &  &  \\
0.735  & $
6.53E-4\pm1.57E-5$ & $195.68\pm4.72$ &   &    &  &  &  &  \\
\hline
+2507  &  &  &  &   &  &  &  &  \\
\hline
0.144  & $
5.17E-4\pm2.56E-6$ & $154.89\pm0.77$ & $5.57E-4$ & $166.90$  &  &  &  &  \\
0.192  & $
5.54E-4\pm2.41E-6$ & $165.95\pm0.72$ & $5.72E-4$ & $171.39$  &  &  &  &  \\
0.243  & $
5.57E-4\pm2.91E-6$ & $166.85\pm0.87$ & $5.77E-4$ & $173.10$  &  &  &  &  \\
0.291  & $
5.77E-4\pm5.51E-6$ & $173.12\pm1.65$ & $5.87E-4$ & $176.10$  &  &  &  &  \\
0.342  & $
6.01E-4\pm2.37E-6$ & $180.12\pm0.71$ & $6.02E-4$ & $180.47$  &  &  &  &  \\
0.391  & $
6.00E-4\pm3.76E-6$ & $179.83\pm1.13$ & $6.17E-4$ & $184.86$  &  &  &  &  \\
\hline
\end{tabular}}
\end{center}
\tablecomments{\\\hspace{\textwidth}
	1. $z_{fit}$ denotes the redshift of H$\alpha$ absorption line, while $V_{rf}$ represents the radial velocity of H$\alpha$ absorption line. $z_{fit}$ and $V_{rf}$ are calculated by the processes of fitting with $S\acute{e}rsic$ function.\\\hspace{\textwidth}
	2. $z_{SLAM}$ denotes the redshift of spectra calculated by SLAM. $V_{rs}$ represents the radial velocity calculated by SLAM. \\\hspace{\textwidth}
	3. $z_{\rm e1,\alpha}$ denotes the redshift of the emission part of the ``first apparition'' in the observer's frame.\\\hspace{\textwidth}
	4. $V_{\rm e1,\alpha}$ denotes the radial velocity of the emission part of the ``first apparition'' in the stellar rest frame.\\\hspace{\textwidth}
	5. Flux$_{\rm e1,\alpha}$ represents the normalized flux of the emission.
	\\\hspace{\textwidth}
	6. FWHM$_{\rm e1,\alpha}$ represents full width at half maximum of the emission.
	\\\hspace{\textwidth}
	7. The measurements with $^*$ indicate low-quality results.} 
\end{table*}

\section{Spectral evolution}\label{sec:specevolution}

\begin{figure*}[tb]
\begin{center}
\begin{tabular}{c}
\includegraphics*[width=180mm,height=15cm]{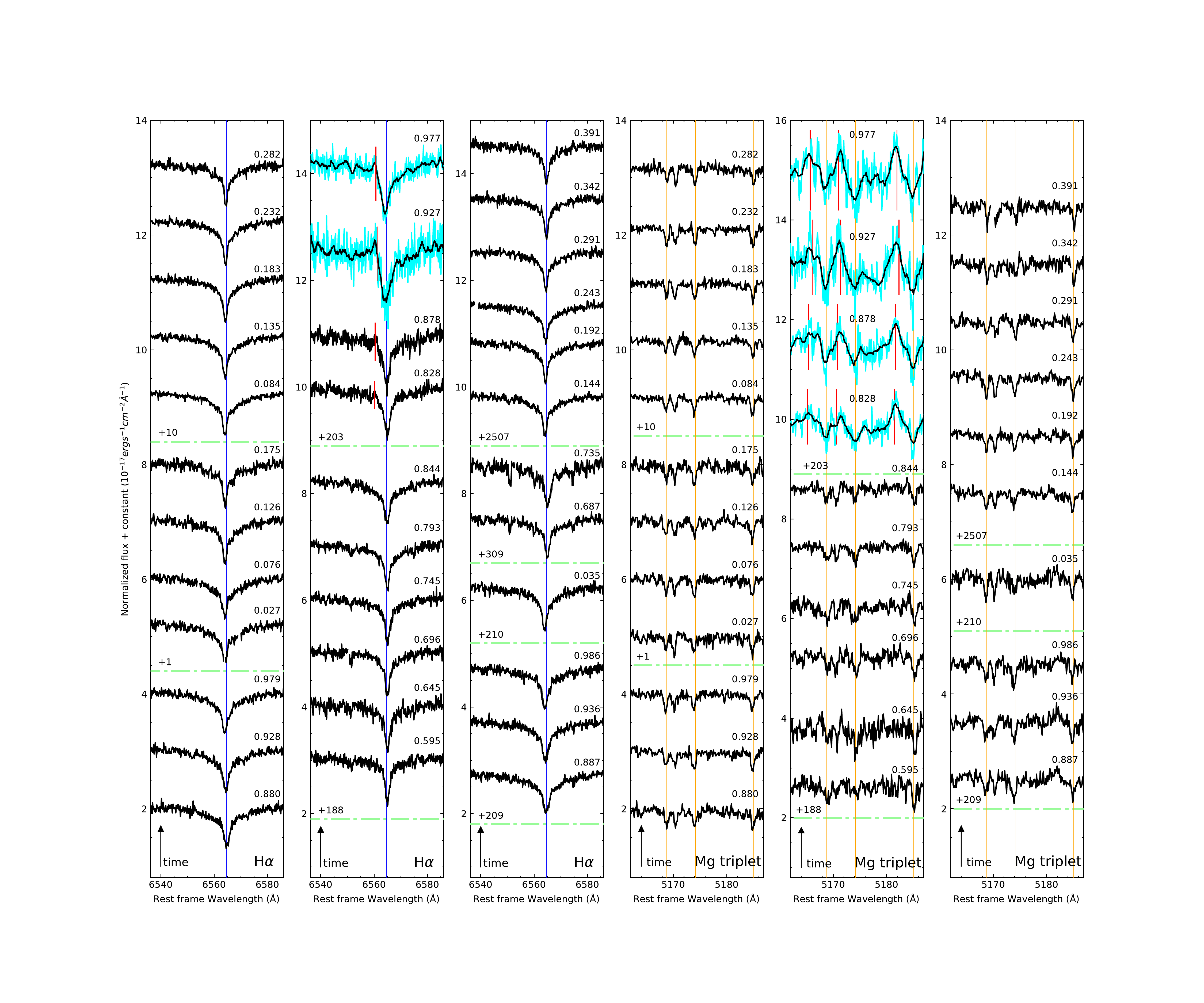}
\end{tabular}
\caption{\label{fig:Halphaevolution} H$\alpha$ and Mg triplet line variations of Catalina-1104058050978 over several pulsation cycles in the system comoving with the pulsation. The spectra are normalized and cosmic rays are cleared. The pulsation cycle of earliest spectrum is considered as the first cycle. The phase is marked at the upper right of the spectrum. Different cycles are divided by green dashed lines. The difference of the number of cycles is marked at the upper left of the green dashed line. Vertical blue line indicates the H$\alpha$ line laboratory wavelength. Vertical orange lines indicate the Mg triplet line laboratory wavelength. For H$\alpha$ lines, in cycle 204, when $\phi = 0.927-0.977$, and for Mg lines, in cycle 204, when $\phi = 0.828-0.977$, the spectra are smoothed. The original normalized spectra are shown in cyan color and the smoothed spectra are marked as black color, while as for other spectra, the original normalized spectra are displayed in black color. Vertical red lines indicate the shifted wavelengths from the H$\alpha$ or Mg absorption lines with the measured blueshifts of H$\alpha$ emission components.}
\end{center}
\end{figure*}

According to the wavelength range, we represent profile variations for H$\alpha$ and Mg lines in Figure~\ref{fig:Halphaevolution}. The spectra are normalized by the fitting results of the continums and shown in the system comoving with the pulsation. The spectra are shifted by the average redshift of the pulsation provided by SLAM. Cosmic rays are removed. The pulsation cycle of earliest spectrum in this time series is set as the cycle 1. It should be noted that the time line is divided as cycle 1$-$2508, so we did not mix up the spectra to one pulsation cycle. The spectroscopic time series displays periodic distortion possibly owing to the turbulent convection \citep{Benk2021MNRAS.500.2554B}.

As for H$\alpha$ lines, in cycle 204, when $\phi = 0.927-0.977$, owing to the large noise, we smooth the spectra. The original normalized spectra is displayed in cyan color and the smoothed spectra are shown in black color. The line asymmetry can be observed over the pulsation cycle. In cycle 204, when $\phi = 0.878-0.977$, a clear blueshifted hydrogen emission is observed. The significances of the emissions shown in Figure~\ref{fig:blueshiftfitting} certify their physical reality. The weather when observing the spectra of cycle 204 was not very good, but there is little prospect that H$\alpha$ and Mg triplet show significant blueshifted emission simultaneously due to accident. It can be explained as the observational characteristic of the main shock before maximum luminosity, called the ``first apparition". In cycle 210, when $\phi = 0.936$, possible weak emission with significance over 2$\sigma$ on the blue wing of H$\alpha$ absorption profile is also shown. In cycle 189, when $\phi = 0.793$, a possible weak blueshifted H$\alpha$ bump with significance over 5$\sigma$ is observed, which can be explained as the ``second apparition''. In cycle 2, when $\phi = 0.175$, a small red bump with significance over 2$\sigma$ is detected, which is possibly corresponding to the ``third apparition''. As for the ``first apparitions'' in cycle 204, the redshift in the frame of the observer $z_{\rm
e1,\alpha}$, the radial velocity in the stellar rest frame $V_{\rm e1,\alpha}$,
the normalized flux Flux$_{\rm e1,\alpha}$ and FWHM of
the emission and absorption FWHM$_{\rm e1,\alpha}$ in the stellar rest frame are measured with two
$S\acute{e}rsic$ functions and displayed in Table~\ref{table:parameterrv}.

As for Mg triplet, in cycle 204, when $\phi = 0.878-0.977$, the ``first apparition'' is observed in H$\alpha$, three emission lines are obviously growing stronger. If the Mg emissions are considered as blueshifted emissions from Mg absorption triplet, they show similar blueshifted velocities as the H$\alpha$ emission in each spectra. This fact indicates that the emissions of H$\alpha$ and Mg triplet are probably from the same physical process.

\section{Discussion}\label{sec:discussion}

\begin{figure*}[tb]
\begin{center}
\begin{tabular}{c}
\includegraphics*[width=120mm,height=8cm]{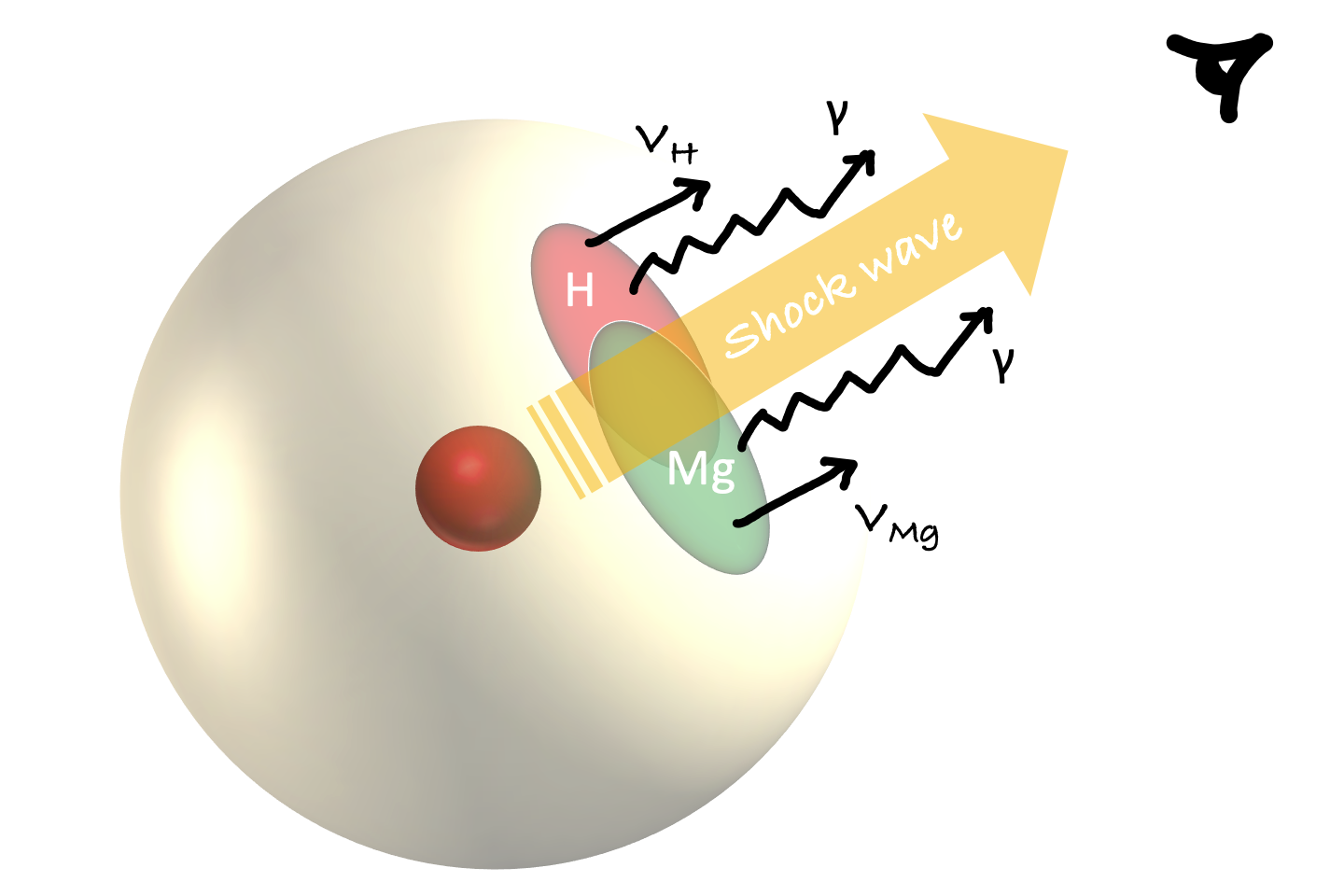}
\end{tabular}
\caption{\label{fig:illustration} Schematic diagram of a probable explanation of the birth process of the blueshifted H$\alpha$ and Mg triplet emission.}
\end{center}
\end{figure*}

We show a probable explanation of the source of the blueshifted emission of H$\alpha$ and Mg triplet in Figure~\ref{fig:illustration}. We present a simple model of a RR Lyrae star. A small dense core is at the center of the RR Lyrae pulsator, which generates energy at a steady rate. The core is wrapped by a large, low-density envelope, which is pulsating periodically. We consider a component in the atmosphere of the RR Lyrae variable which contains H and Mg. This component is not only consisting of H and Mg. In this work we only present the spectroscopic observations of H$\alpha$ and Mg lines. So only H and Mg are emphasized in the illustration. When the shock wave is travelling through this component, the atoms are excited. After the hot shock front passed, the atoms de-excite in the cooler radiative wake, with the redundant energy released. The blueshift of the emission indicates the motion of the shock wave. The similar radial velocity of the blueshifted emission of H$\alpha$ and Mg suggests that they are probably triggered by the same shock wave process. This further confirms the existence of shock wave in non-fundamental mode RR Lyrae pulsators after \cite{Duan2021ApJ...909...25D}.

\cite{Chadid2011Carnegie} reported that the blueshifted hydrogen emssion presents over $\sim 5\%$ of the whole period. \cite{Gillet2019} showed that the ``first apparition'' appears during $\phi=0.892-0.929$ in RR Lyr, accounting for about $3.7\%$ of the pulsation cycle. In this work, the blueshifted hydrogen emission in cycle 204 from $\phi = 0.828$ to $0.977$ accounts $\sim 15\%$ for the whole pulsation cycle. We suggest that in some RR Lyrae pulsators, the shock-triggered emission may last much longer than the previous observations. Moreover, the pulsator does not show equal intense emission before the maximum luminosity in every cycle. The shock process is unstable in this pulsator. Possible weak blueshifted bump with significance over 3$\sigma$ can also be observed in cycle 2508, when $\phi = 0.342-0.391$, which indicates a possible weak shock process besides the ``three apparitions''.

\section{Conclusions}\label{sec:concl}

In this work, we present spectroscopic and photometric time series analyses of a non-Blazhko RRc pulsator, Catalina-1104058050978, with LAMOST medium resolution spectra and Catalina Sky Survey. Blueshifted magnesium triplet emissions are reported.

We analyse the photometric observations from Catalina sky survey of this RRc pulsator with pre-whitening sequence method. An additional frequency signal $f_x = 4.34579$ with $P_1/P_x = 0.69841$ is detected and discussed. According to the regenerated light curve, we determine the ephemeris and phases of Catalina-1104058050978. The redshift and radial velocity curve of the spectra are provided by fitting process and SLAM.

Moreover, the variation of H$\alpha$ and Mg lines in the system comoving with the pulsation is displayed. Significant evolution of blueshifted hydrogen emission and Mg emission possibly related to the main shock process are observed, which further confirms the existence of shock wave in non-fundamental mode RR Lyrae pulsators. Radial velocities of absorptions and emissions of H$\alpha$ lines and parameters of the blueshifted emissions are measured and displayed in this work. The blueshifted hydrogen emission in Catalina-1104058050978 lasts $\sim 15\%$ of the whole pulsation cycle. We suggest that the duration of shock-triggered emission can be much longer than the previous observations in some RR Lyrae pulsators.

The atmospheric dynamics of non-fundamental RR Lyrae pulsators is still unclear owing to the limited observations. Now we can determine that shock wave is not absent in the atmosphere of RRc pulsators. With
more upcoming time series, the physical mechanism of RRc pulsators will be unveiled.

\section*{Acknowledgements}\label{sec:ackno}

Xiao-Wei Duan acknowledges research support from the
Cultivation Project for LAMOST Scientific Payoff and Research Achievement of
CAMS-CAS and the Lee Wai Wing Scholarship. Li-Cai Deng thanks research support from the National Science
Foundation of China through grants 11633005. Xiao-Dian Chen also thanks
support from the National Natural Science Foundation of China through grant 
11903045. Xiao-Dian Chen and Xiao-Wei Duan thank the support from the National Natural Science Foundation of China through grant 12173047. Hua-Wei Zhang thanks research support from the
National Natural Science Foundation of China (NSFC) under No. 11973001 and
National Key R$\&$D Program of China No. 2019YFA0405504. We acknowledge Mark Taylor
for the TOPCAT software. 
The Large Sky Area Multi-Object Fiber Spectroscopic
Telescope (Guoshoujing Telescope) is a National Major Scientific Project built by the Chinese
Academy of Sciences, funded by the National
Development and Reform Commission. LAMOST is operated and managed by the
National Astronomical Observatories, Chinese Academy of Sciences. 
The Catalina Sky Survey survey is supported by the National Aeronautics and Space Administration
under Grant No. NNG05GF22G.  
The CRTS survey is supported by the
U.S.~National Science Foundation under grants AST-0909182 and AST-1313422.

\vspace{5mm} 
\software{NumPy \citep{Harris2020Natur.585..357H}, SciPy
\citep{Virtanen2020SciPy-NMeth}, AstroPy
\citep{Astropyprice2018astropy}, Matplotlib
\citep{Matplotlibhunter2007matplotlib},
Scikit-learn \citep{Scikitpedregosa2011scikit}, 
PyAstronomy \citep{Czesla2019ascl.soft06010C}, TOPCAT
\citep{Taylor2005ASPC..347...29T}}

\bibliographystyle{aasjournal}  
\bibliography{mainDXWnewMg}


\end{document}